\documentclass[]{IEEEtran}

\usepackage{cite}
\usepackage{amsmath,amssymb,amsfonts}
\usepackage{algorithmic}
\usepackage{graphicx}
\usepackage{textcomp}
\usepackage{xcolor}
\usepackage{subfig}

\usepackage{bm}
\usepackage{xspace}
\newcommand{\bmath}[1]{\ensuremath{\bm{#1}}\xspace}

\newcommand{\x}{\bmath{x}}
\newcommand{\y}{\bmath{y}}
\newcommand{\z}{\bmath{z}}

\newcommand{\f}{\bmath{f}}

\newcommand{\n}{\bmath{n}}

\newcommand{\uv}{\bmath{u}}

\newcommand{\rv}{\bmath{r}}
\newcommand{\w}{\bmath{w}}

\newcommand{\alp}{\bmath{\alpha}}

\newcommand{\tht}{\bmath{\theta}}
\newcommand{\lam}{\bmath{\lambda}}

\newcommand{\ro}{\bmath{\rho}}

\newcommand{\muv}{\bmath{\mu}}

\newcommand{\phiv}{\bmath{\phi}}

\newcommand{\A}{\bmath{A}}

\newcommand{\G}{\bmath{G}}

\newcommand{\K}{\bmath{K}}

\newcommand{\U}{\bmath{U}}

\newcommand{\beq}{\begin{equation}}
\newcommand{\eeq}{\end{equation}}
\newcommand{\bea}{\begin{eqnarray}}
\newcommand{\eea}{\end{eqnarray}}
\newcommand{\ba}{\left(\!\!\begin{array}}
\newcommand{\ea}{\end{array}\!\!\right)}
\newcommand{\bc}{\begin{center}}
\newcommand{\ec}{\end{center}}

\newcommand{\diag}{\mathrm{diag}}

\def\BibTeX{{\rm B\kern-.05em{\sc i\kern-.025em b}\kern-.08em
		T\kern-.1667em\lower.7ex\hbox{E}\kern-.125emX}}
\newcommand{\ov}{\bmath{o}}
\newcommand{\cnn}{\mathrm{CNN}}
\newcommand{\patch}{\mathrm{Patch}}
\newcommand{\nd}{{N_\mathrm{d}}}
\newcommand{\nj}{{N_\mathrm{p}}}
\newcommand{\nt}{{N_\mathrm{t}}}
\newcommand{\nr}{{N_\mathrm{r}}}

\newcommand{\txtc}[1]{\textcolor{black}{#1}}
\newcommand{\txtr}[1]{\textcolor{black}{#1}}
\newcommand{\txtb}[1]{\textcolor{blue}{#1}}

\begin{document}

\title{Modified Kernel MLAA Using Autoencoder for PET-enabled Dual-Energy CT
	\thanks{This work is supported in part by National Institutes of Health (NIH) under the grant no. R21EB027346. 
	Email: sqlli@ucdavis.edu, gbwang@ucdavis.edu.}
}

\author{\IEEEauthorblockN{Siqi Li and Guobao Wang}
	
	\IEEEauthorblockA{University of California Davis Medical Center}
}

\maketitle

\begin{abstract}
Combined use of PET and dual-energy CT provides complementary information for multi-parametric imaging.  PET-enabled dual-energy CT combines a low-energy x-ray CT image with a high-energy $\gamma$-ray CT (GCT) image reconstructed from time-of-flight PET emission data to enable dual-energy CT material decomposition on a PET/CT scanner.  The maximum-likelihood attenuation and activity (MLAA) algorithm has been used for GCT reconstruction but suffers from noise. Kernel MLAA exploits an x-ray CT image prior through the kernel framework to guide GCT reconstruction and has demonstrated substantial improvements in noise suppression. However, similar to other kernel methods for image reconstruction, the existing kernel MLAA uses image intensity-based features to construct the kernel representation, which is not always robust and may lead to suboptimal reconstruction with artifacts. In this paper, we propose a modified kernel method by using an autoencoder convolutional neural network (CNN) to extract an intrinsic feature set from the x-ray CT image prior.  A computer simulation study was conducted to compare the autoencoder CNN-derived feature representation with raw image patches for evaluation of kernel MLAA for GCT image reconstruction and dual-energy multi-material decomposition. 
The results show that the autoencoder kernel MLAA method can achieve a significant image quality improvement for GCT and material decomposition as compared to the existing kernel MLAA algorithm. A weakness of the proposed method is its potential over-smoothness in a bone region, indicating the importance of further optimization in future work.
The codes is available on \txtb{\emph{https://github.com/SiqiLi1020/Autoencoder-Kernel-MLAA}}.
\end{abstract}



\section{Introduction}
Positron emission tomography (PET) integrated with computed tomography (CT) is a molecular imaging modality that is widely used in clinical oncology, neurology and cardiology. In parallel, dual-energy (DE) CT imaging has the unique capability of using energy-dependent tissue attenuation information to perform quantitative multi-material decomposition \cite{b7}. Combined use of PET/CT and DECT provides a multi-parametric characterization of disease states in cancer and other diseases \cite{b23}. Nevertheless, the integration of DECT with existing PET/CT would not be trivial, either requiring costly CT hardware upgrade or significantly increasing CT radiation dose.

We have proposed a new dual-energy CT imaging method that is enabled using a standard time-of-flight PET/CT scan without change of scanner hardware or adding additional radiation dose or scan time \cite{b12, Wang20}. Instead of using two different x-ray energies as commonly used by conventional DECT, the PET-enabled dual-energy CT method combines a radiotracer annihilation-generated high-energy ``$\gamma$-ray CT (GCT)" at 511 keV with the already-available low-energy x-ray CT (usually $\leq140$ keV) to produce a pair of dual-energy CT images on PET/CT for multi-material decomposition. 

The reconstruction of GCT image from the PET emission scan can be achieved by the maximum likelihood attenuation and activity (MLAA) method \cite{b8}. 
However, standard MLAA reconstruction is commonly noisy because the counting statistics of PET emission data is limited. While the noise would not compromise the performance of MLAA for PET attenuation correction, it may affect the quantitative accuracy of GCT for multi-material decomposition. To suppress noise, the kernel MLAA approach \cite{Wang20} has been developed by use of x-ray CT as image prior through a kernel framework and has demonstrated substantial improvements over standard MLAA. 

In the kernel methods for image reconstruction (e.g. \cite{Wang2015, Hutchcroft16, Novosad16, Gong18, Deidda19}), a set of features need to be defined for constructing the kernel representation of the image to be estimated. Existing kernel methods have mainly used image pixel intensities of a small patch (e.g., for MR-guided PET reconstruction \cite{Hutchcroft16, Novosad16, Gong18, Deidda19}) or temporal sequence (e.g., for dynamic PET reconstruction \cite{Wang2015, Gong18}). However, the intensity-based features do not always provide satisfactory results.
As shown in \cite{Wang20} and later in this paper, the reconstructed GCT image by such a method suffers from artifacts. 

In this paper, we propose to use a convolutional neural network (CNN) feature set that is adaptively learned on the prior image to build the kernel representation for MLAA reconstruction. It has been demonstrated that deep learning with CNN has a strong ability to derive a latent feature representation in different tasks \cite{b18}. While it is often impractical to collect a large amount of training data for supervised learning, here we utilize the concept of autoencoder CNN \cite{b19}, an unsupervised representation learning technique, for intrinsic feature extraction from the x-ray CT image prior for the kernel construction. 
The autoencoder CNN-derived feature set of x-ray CT image is expected to provide a more robust kernel representation for the GCT image reconstruction than the conventional intensity-based features.

\section{PET-enabled Dual-Energy CT}

\subsection{Statistical Model of PET Emission Data}
In time-of-flight (TOF) PET, the measured data $\y$ can be well modeled as independent Poisson random variables using the log-likelihood function,
\beq
L(\y|\lam, \muv) = \sum_{i=1}^\nd\sum_{m=1}^\nt y_{i,m}\log\overline{y}_{i,m}(\lam,\muv)-\overline{y}_{i,m}(\lam,\muv),
\eeq
where $i$ denotes the index of PET detector pair and $\nd$ is the total number of detector pairs. $m$ denotes the $m$th TOF bin and $\nt$ is the number of TOF bins. 
The expectation of the PET projection data $\overline{\y}$ is related to the radiotracer activity image \txtc{$\lam\in R^{\nj\times 1}$ and object attenuation image $\muv\in R^{\nj\times 1}$} at 511 keV via
\beq
\overline{\y}_{m}(\lam, \muv) = \diag\{\n_m(\muv)\}\G_m\lam + \rv_m,
\label{1}
\eeq
where \txtc{$\G_m\in R^{\nd\times\nj}$} is the PET detection probability matrix and $g_{ij}$ is the probability of detecting an event originated in pixel $j$ by detector pair $i$. \txtc{$\nj$ is the total number of image pixels.} $\rv_m$\txtc{$\in R^{\nd \times 1}$} accounts for the expectation of random and scattered events. $\n_m(\muv)$\txtc{$\in R^{\nd \times 1}$} is the normalization factor with the $i$th element being
\beq
n_{i,m}(\muv) = c_{i,m} \cdot \exp(-[\A\muv]_i),
\eeq
where $c_{i,m}$ represents the multiplicative factor excluding the attenuation correction factor and $\A$\txtc{$\in R^{\nd \times \nj}$} is the system matrix for transmission imaging.

\begin{figure*}[t]
	\centering
	\subfloat[RED-CNN]{\includegraphics[trim=1cm 13cm 1cm 0.0cm, clip, width=15cm]{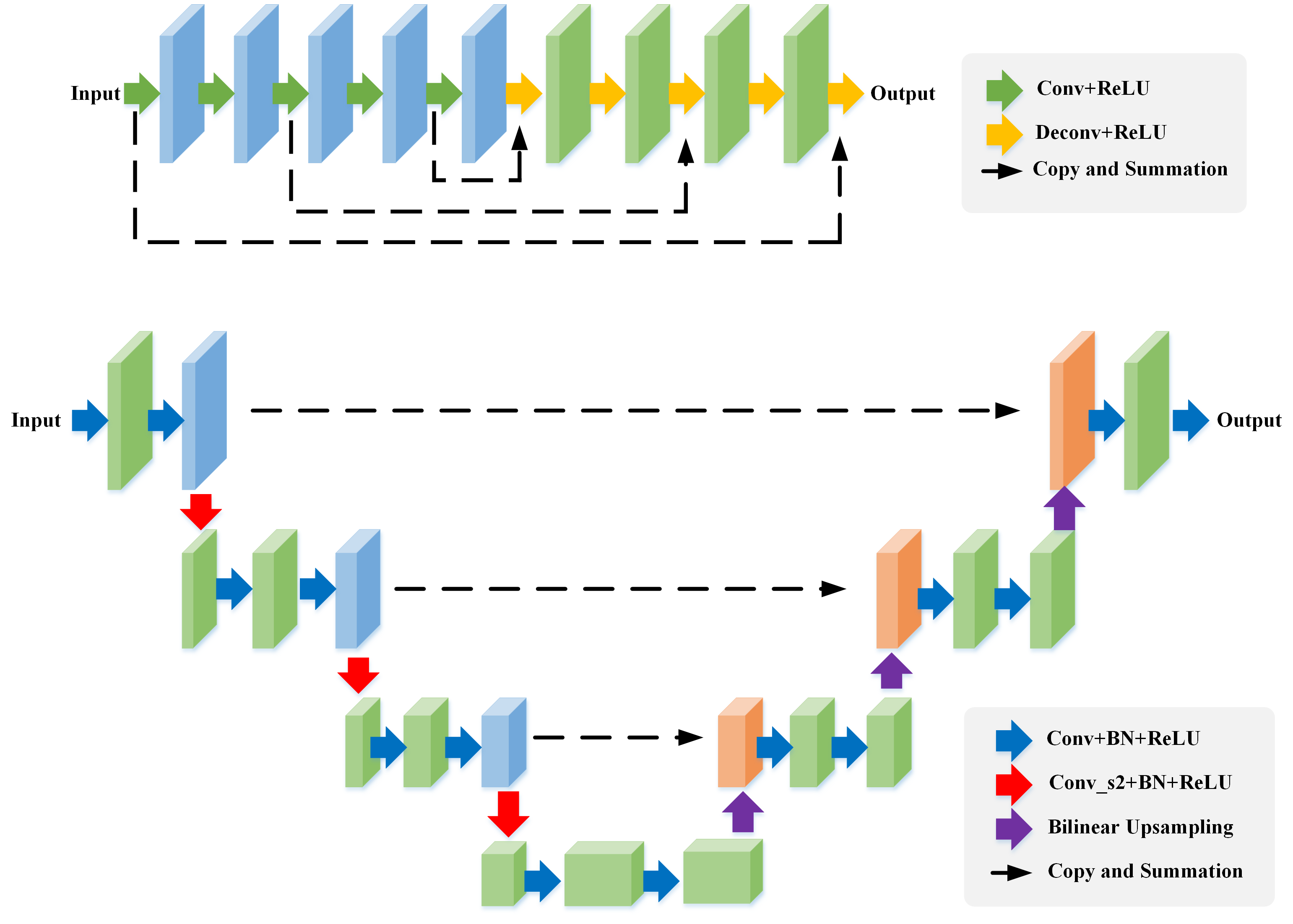}}\\
	\subfloat[Unet]{\includegraphics[trim=0cm 0cm 0cm 6cm, clip, width=15cm]{red_and_unet.png}}
	\caption{CNN models used for feature extraction in an unsupervised form. (a) RED-CNN. (b) Unet. \txtc{Each solid arrow denotes a layer and each 3D block indicates the feature output of a layer.}}
	\label{fig-1}
\end{figure*}
\begin{figure*}[t]
	\vspace{-8pt}
	\centering
	\subfloat[]{\includegraphics[trim=2cm 1cm 1cm 0.0cm, clip, width=4.42cm]{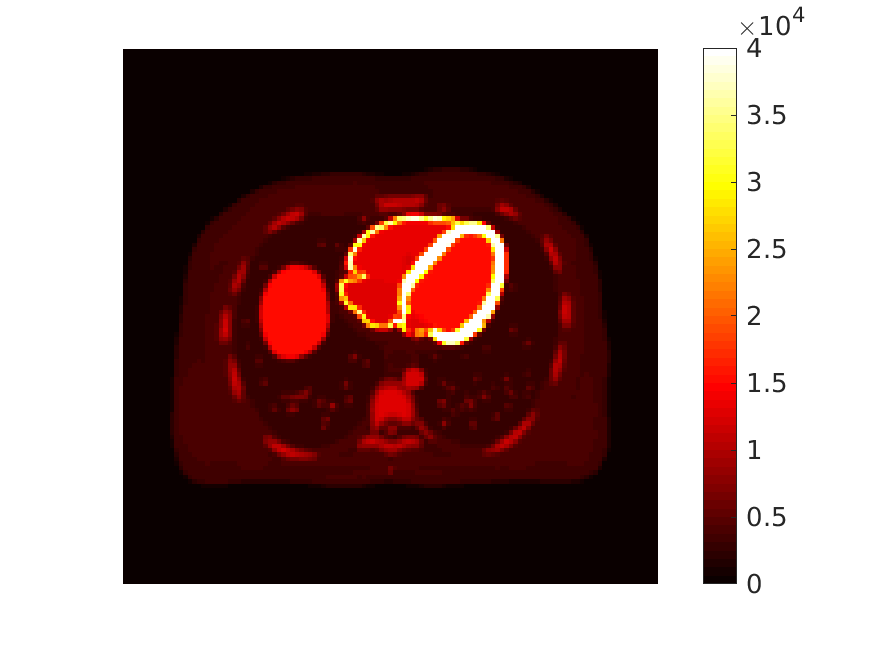}}
	\subfloat[]{\includegraphics[trim=2cm 1cm 1cm 0.5cm, clip, width=4.5cm]{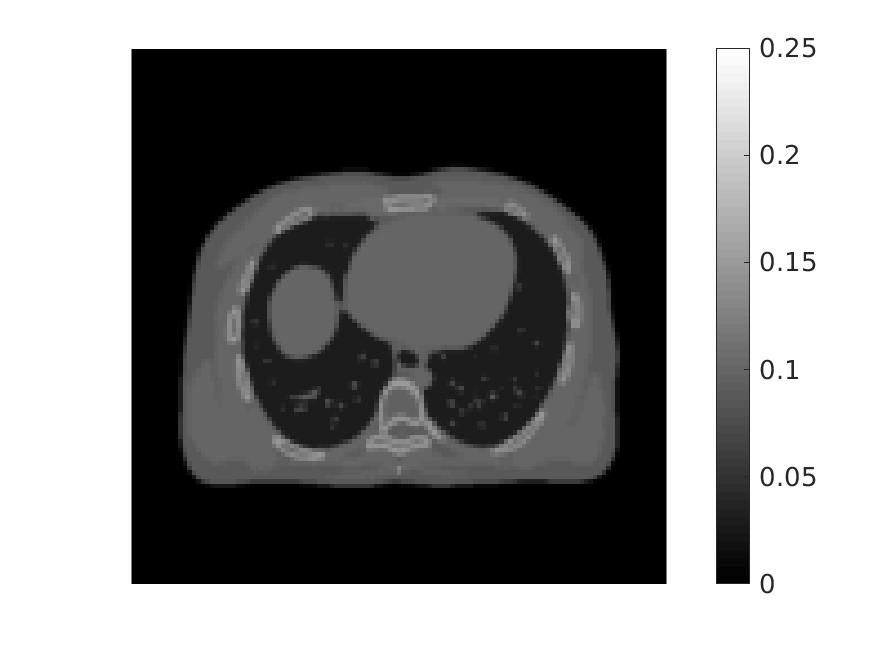}}
	\subfloat[]{\includegraphics[trim=2cm 1cm 1cm 0.5cm, clip, width=4.5cm]{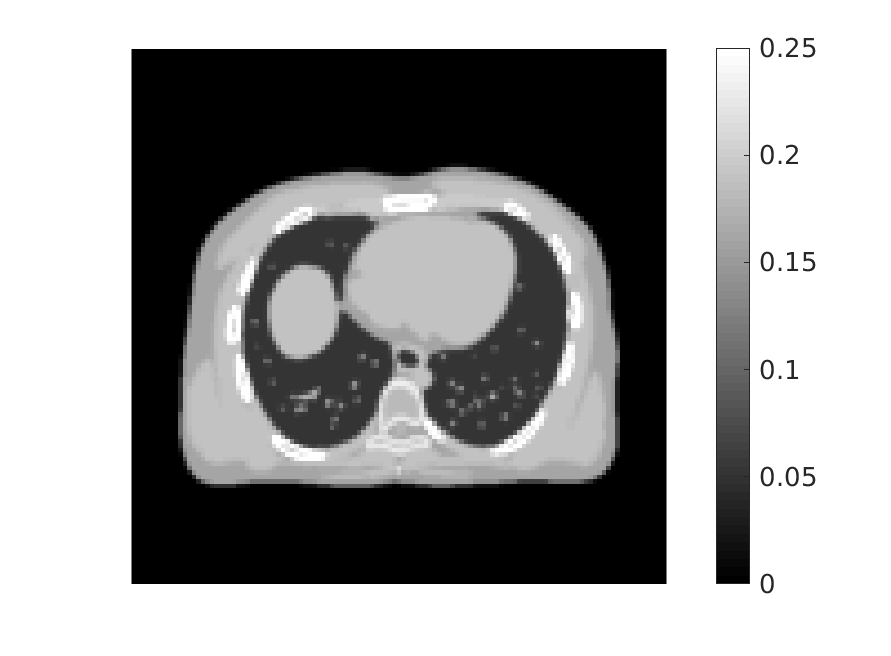}}
	\subfloat[]{\includegraphics[trim=2cm 1.6cm 1cm 0.5cm, clip, width=4.5cm]{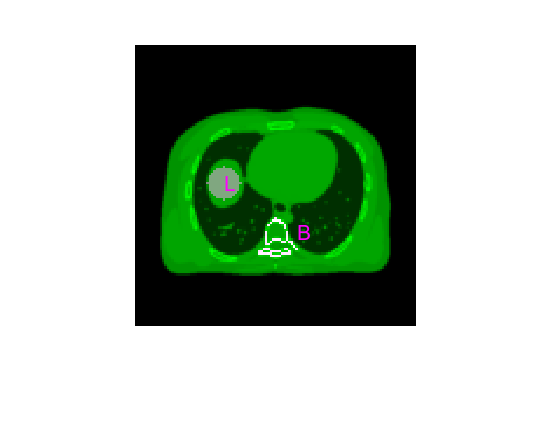}}
	\caption{The digital phantom used in the PET/CT computer simulation. (a) PET activity image in Bq/cc; (b) PET attenuation image at 511 keV in cm$^{-1}$; (c) x-ray CT image at 80 keV; (d) illustration of a liver ROI `L' and a spine bone ROI `B'.}
	\label{fig-pht-xcat}\vspace{-0pt}
\end{figure*}
\subsection{Maximum Likelihood Attenuation and Activity (MLAA) Reconstruction}

The maximum-likelihood attenuation and activity (MLAA, \cite{b8}) reconstruction jointly estimates the attenuation image $\muv$ and the activity image $\lam$ from the projection data $\y$ by maximizing the Poisson \txtc{log-likelihood},
\beq
\hat{\lam}, \hat{\muv} = \txtc{\arg\max}_{\lam \geq 0, \muv \geq 0}L(\y|\lam,\muv).
\label{eq-mlaa}
\eeq
An iterative interleaved updating strategy is commonly used to seek the solution \cite{b8}. 

In each iteration of the algorithm, $\lam$ is first updated with fixed attenuation image $\hat{\muv}$,
\beq
\hat{\lam} = {\arg\max}_{\lam \geq 0} L(\y|\lam,\hat{\muv}),
\eeq
which can be solved by the maximum-likelihood expectation maximization (MLEM) algorithm \cite{b3}.
$\muv$ is then updated with fixed $\lam$ using the maximum-likelihood transmission reconstruction (MLTR) method,
\beq
\hat{\muv} = {\arg\max}_{\txtc{\muv} \geq 0} L(\y|\hat{\lam},\muv),
\eeq
which can be solved by the separable paraboloidal surrogate (SPS) algorithm \cite{b4}. 

Previous use of MLAA was mainly for improving PET attenuation correction (e.g.,  \cite{b25,b26}) or transmission-less PET imaging (e.g., \cite{b10,b11,b29,b32}). In our PET-enabled dual-energy CT method \cite{b12, Wang20}, the MLAA is used differently. The estimated high-energy GCT image $\muv$ is combined with the low-energy x-ray CT image $\x$ to form dual-energy imaging for multi-material decomposition.

\subsection{Kernel MLAA}

The GCT estimate by  standard MLAA is commonly noisy due to the limited counting statistics of PET emission data. To suppress noise, the kernel MLAA approach  \cite{Wang20} incorporates the x-ray CT image as \emph{a priori} information to guide the GCT reconstruction in the MLAA. It describes the intensity of the GCT $\mu_j$ in pixel $j$ as a linear representation in a transformed feature space 
\beq
\mu_j = \w^{T}\phiv(\f_j),
\eeq
where $\f_j$ is the data point of pixel $j$ that is extracted from $\x$ and $\phiv(\f_j)$ is a mapping function that transforms the low-dimensional data point $\f_j$ to a high-dimensional feature vector. $\w$ is a weight vector which also sits in the transformed space,
$
\w = \sum_{l}\alpha_l\phiv(\f_l),
$
with $\alpha$ being the coefficient vector. Then, we can obtain the following kernel representation for $\mu_j$,
\beq
\mu_j = \sum_{l}\alpha_l \phiv(\f_j)^T\phiv(\f_l) = \sum_{l}\alpha_l \kappa(\f_j, \f_l),
\eeq
where $\kappa (\cdot, \cdot)$ is the kernel function (e.g., radial Gaussian) that is equal to the inner product of the two transformed feature vectors $\phiv(\f_j)$ and $\phiv(\f_l)$. 
The equivalent matrix-vector form for the GCT image is 
\beq
\muv = \K\alp,
\label{eq-ker}
\eeq
where $\K$ is the kernel matrix and $\alp$ denotes the corresponding kernel coefficient image. 

Substituting Eq. (\ref{eq-ker}) into the MLAA formulation in Eq. (\ref{eq-mlaa}) gives the following kernel MLAA optimization formulation,
\beq
\hat{\lam},\hat{\alp} = {\arg\max}_{\lam \geq 0, \alp \geq 0}L\big(\y |\lam, \K\alp)\big).
\eeq
The detail of the kernel MLAA algorithm is provided in \cite{Wang20}.
Once $\hat{\alp}$ is obtained, the final estimate of the GCT image is obtained by
$
\hat{\muv} = \K \hat{\alp}.
$

\subsection{Material Decomposition Using PET-enabled Dual-Energy CT}

For each image pixel $j$, the GCT attenuation value $\mu_j$ and x-ray CT attenuation value $x_j$ jointly form a pair of dual-energy measurements $\uv_j\triangleq [x_j, \mu_j]^T$, which can be modeled by a set of material bases, such as air (A), soft tissue (S) or equivalently water, and bone (B):
\beq
\uv_j  = \U\ro_j,  \quad\U\triangleq\left( \begin{array}{ccc}
	x_A & x_S & x_B \\
	\mu_A & \mu_S & \mu_B
\end{array} \right), \ro_j\triangleq\left( \begin{array}{c}
	\rho_{j,A}\\
	\rho_{j,S}\\
	\rho_{j,B}
\end{array} \right),
\eeq
subject to
$
\sum_{k}\rho_{j,k} = 1.
$
The coefficients $\rho_{j,k}$ with $k = {A, S, B}$ are the fraction of each basis material in pixel $j$. The material basis matrix $\U$ consists of the linear attenuation coefficients of each basis material measured at the low and high energies. Finally, $\ro_j$ is estimated using the following least-square optimization for each image pixel,
\beq
\hat{\ro}_j =\arg\max_{\ro_j}\left \|\uv_j - \U\ro_j \right\|.
\eeq
\begin{figure*}[t]
	\vspace{-10pt}
	\centering
	\subfloat[]{\includegraphics[trim=4.3cm 1cm 0.5cm 0.5cm, clip, height=6.4cm]{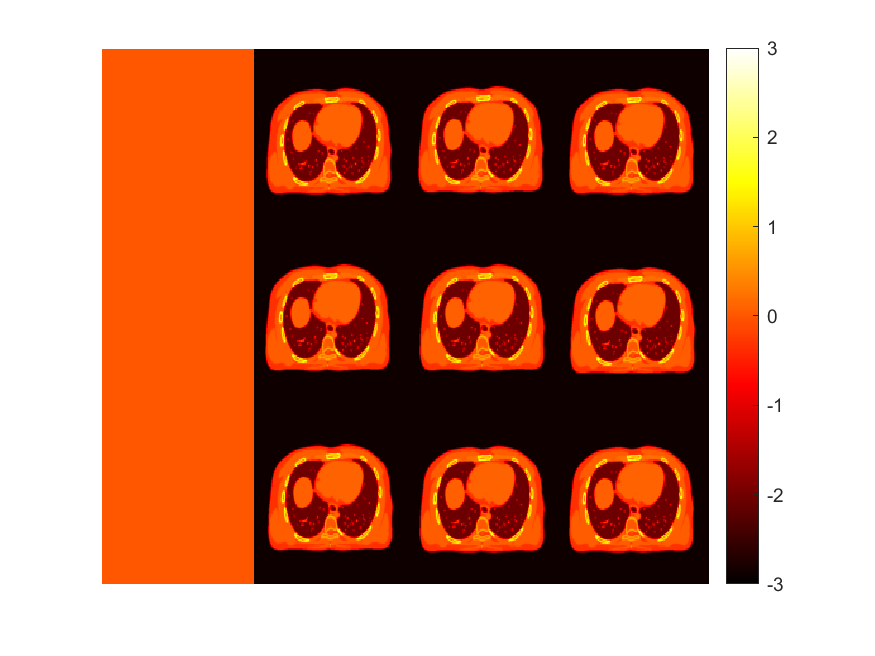}}
	\subfloat[]{\includegraphics[trim=1.5cm 1cm 1.5cm 0.5cm, clip, height=6.4cm]{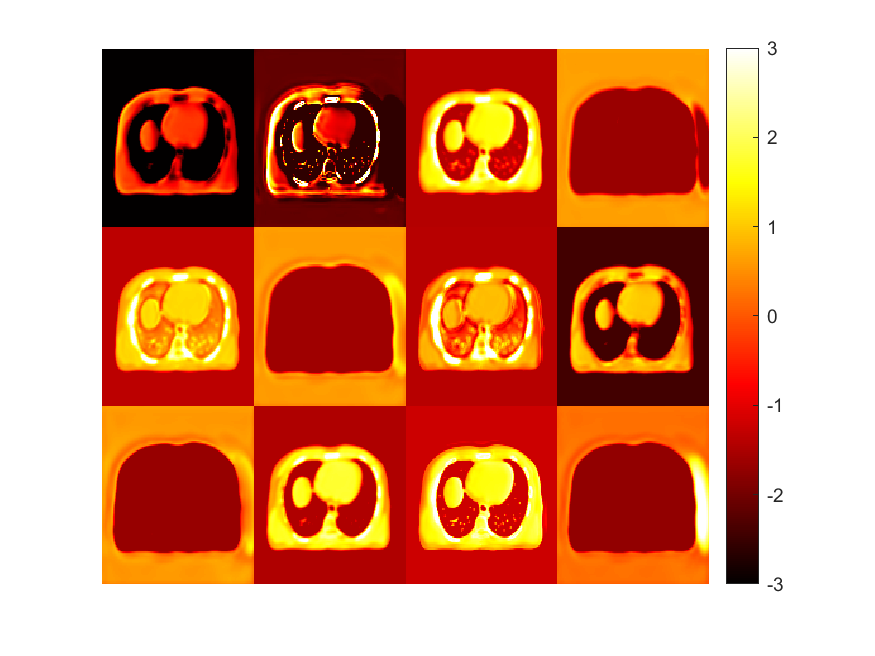}}
	\caption{Maps of the feature set used by (a) standard kernel and (b) proposed Unet kernel.}
	\label{fig-10}
\end{figure*}

\begin{figure*}[t]
	\vspace{-10pt}
	\centering
	\subfloat[]{\includegraphics[trim=1cm 0cm 0.5cm 0cm, clip, height=6.3cm]{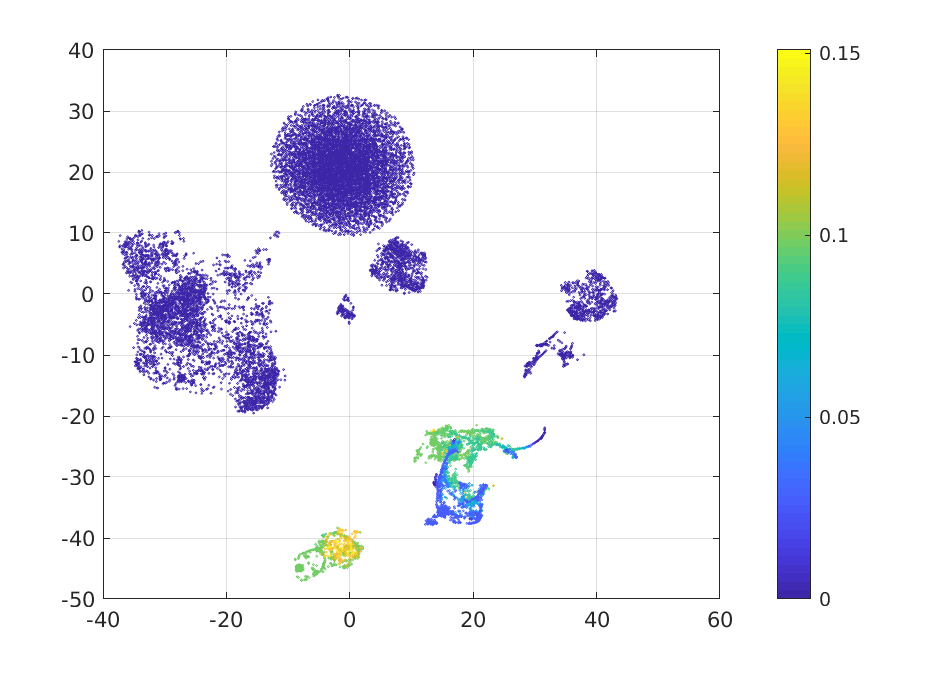}}
	\subfloat[]{\includegraphics[trim=1cm 0cm 0.5cm 0cm, clip, height=6.3cm]{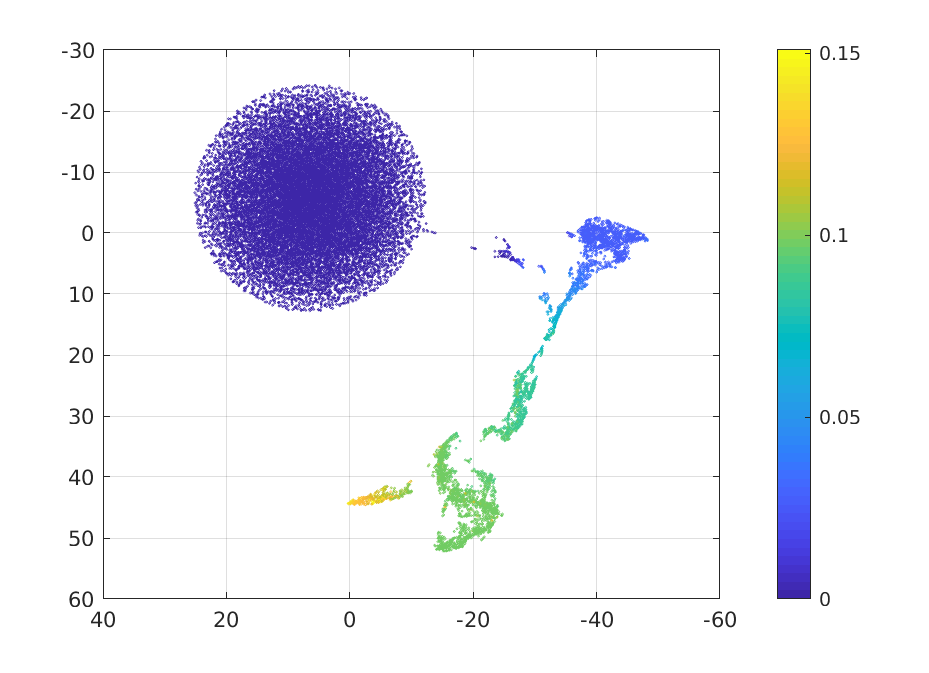}}
	\caption{\txtc{Low-dimensional manifold visualization of (a) standard intensity-based and (b) CNN-derived feature sets. Each axis represents \txtr{a dimension} of the manifold embedding.  Each point corresponds to an image pixel and color represents different GCT attenuation values.}}
	\label{fig-tsne}
\end{figure*}
\section{Proposed Autoencoder Kernel Method}

\subsection{Building Kernels Using CNN Features}

In the kernel MLAA \cite{Wang20} and other kernel methods for image reconstruction (e.g. \cite{Wang2015, Hutchcroft16, Novosad16, Gong18, Deidda19}), the formation of the pixel-wise feature vector $\f_j$ is a key factor to build the kernel matrix $\K$ and directly impacts the reconstruction result. Conventionally, the feature vector $\f_j$ is defined by the intensity value of a pixel or its surrounding small patch (therefore denoted by $\f_j^\patch$). However, such an approach may lead to suboptimal feature representation because of the simplification of feature attributes and the small size of receptive field. Artifacts were observed in the reconstructed GCT images of the kernel MLAA \cite{Wang20}. While it is possible to design more complex features (e.g. texture) to make the features more robust and differential, it would require a significant amount of \txtc{handcrafting}.

To alleviate the issues, we propose to exploit deep learning with convolutional neural networks (CNN) \cite{b18} for intrinsic feature extraction from the x-ray CT image instead of using conventional intensity-based features in the kernel MLAA. The latent feature representation of a trained CNN model may benefit from the larger receptive field of deep convolutions to build a deeper and more robust feature set for kernel representation. A natural choice is by use of supervised deep learning which has shown a strong potential for feature extraction in image recognition tasks. The optimization problem can be accordingly formulated as

\beq
\hat{\tht}=\arg\min_{\tht} \sum_i||\ov_i-\phi(\tht;\z_i)||^2,
\label{eq-sdl}
\eeq
where $i$ denotes the index of training pairs with $\z_i$ the input and $\ov_i$ the output. $\phi(\tht;\cdot)$ is a neural-network model with $\tht$ the unknown weights. Once the neural-network model is pre-trained, we can apply it to the x-ray CT image prior $\x$ to extract intermediate CNN feature sets for each image pixel.

One major challenge with supervised deep learning is it commonly requires a large number of training data sets, which are not always available or the data acquisition is costly.  Therefore, here we explore the feasibility of feature extraction using unsupervised deep learning for the kernel methods.

\begin{figure*}[t]
	\vspace{-10pt}
	\centering
	\subfloat[]{\includegraphics[trim=2cm 1cm 2.8cm 0.0cm, clip, height=5.2cm]{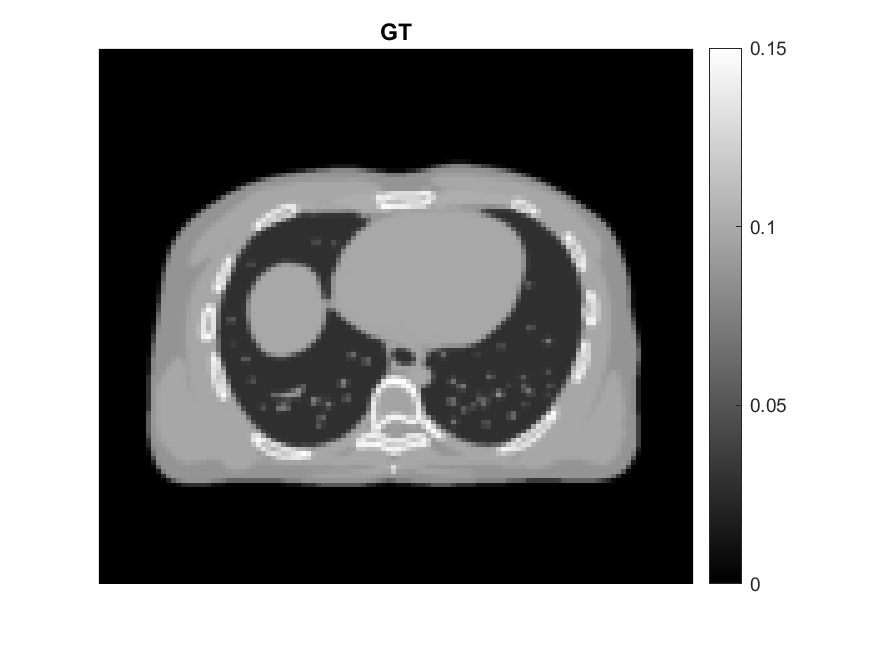}}
	\subfloat[]{\includegraphics[trim=2cm 1cm 2.8cm 0.0cm, clip, height=5.2cm]{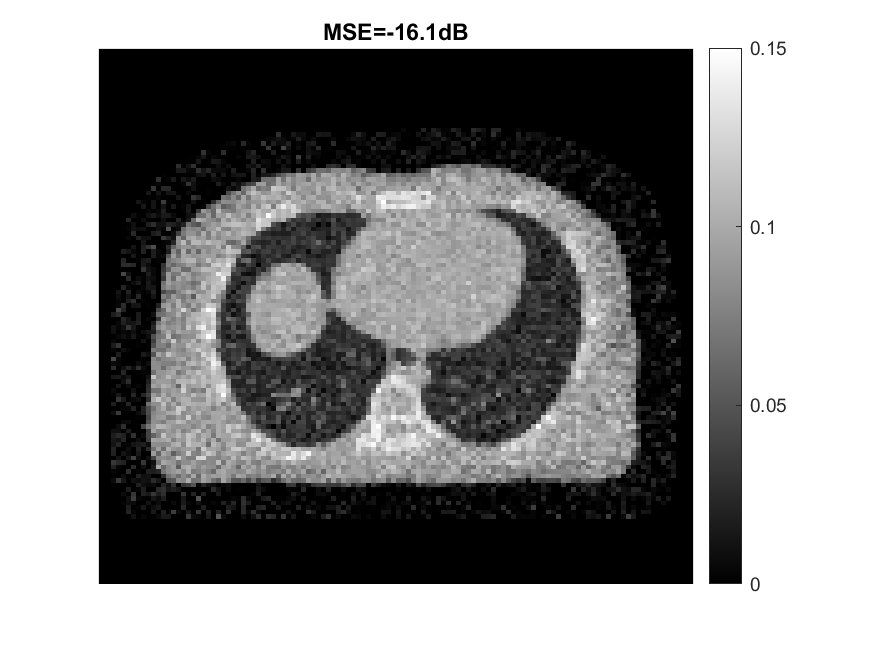}}
	\subfloat[]{\includegraphics[trim=2cm 1cm 1cm 0.0cm, clip, height=5.2cm]{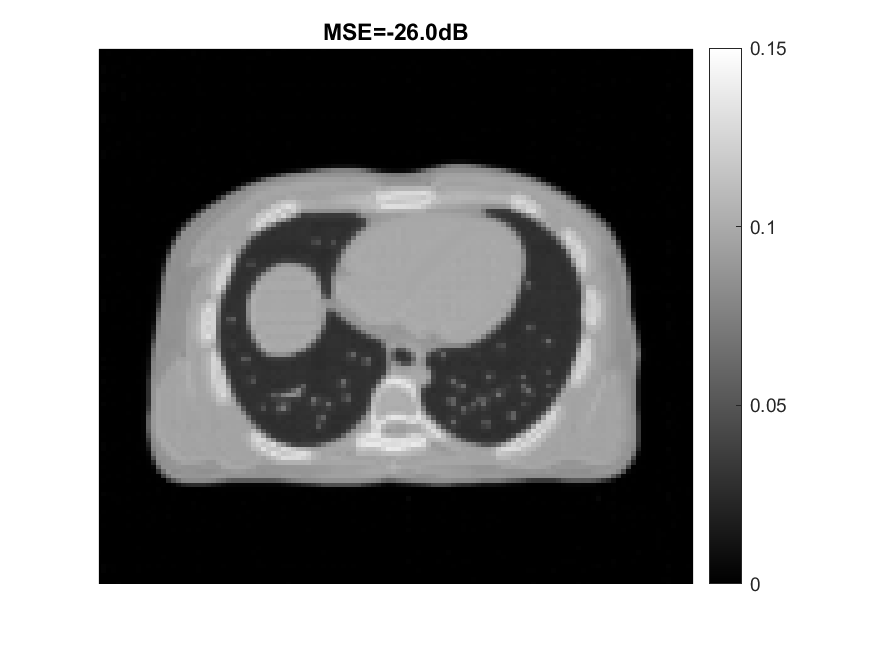}}\\
	\subfloat[]{\includegraphics[trim=2cm 1cm 2.8cm 0.0cm, clip, height=5.2cm]{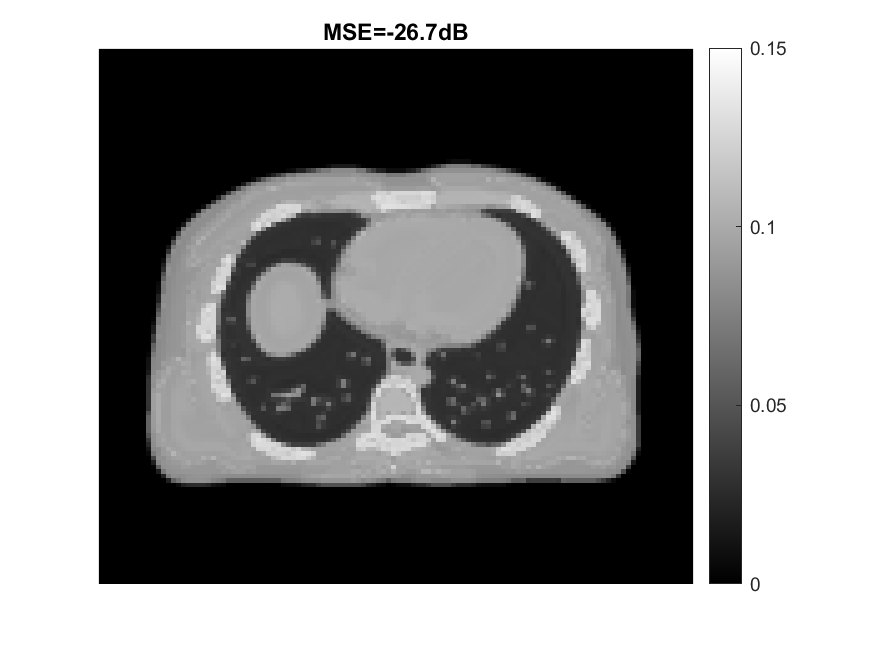}}
	\subfloat[]{\includegraphics[trim=2cm 1cm 1cm 0.0cm, clip, height=5.2cm]{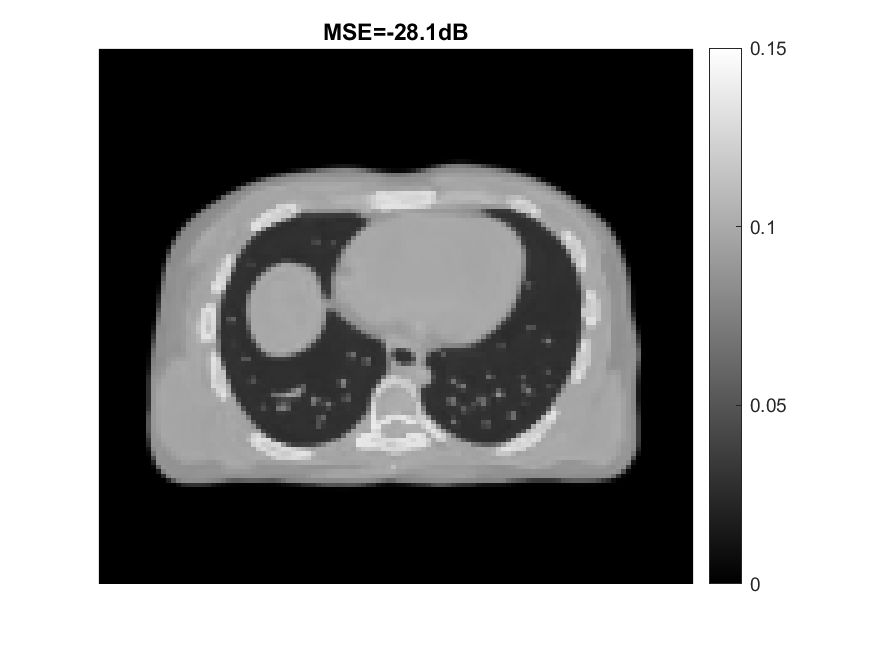}}
	\caption{GCT images by different reconstruction algorithms. (a) Ground truth, (b) standard MLAA, (c) standard kernel MLAA, (d) RED-CNN kernel MLAA, and (e) Unet kernel MLAA.}
	\label{fig-4}
\end{figure*}
\subsection{The Autoencoder Kernel}

An autoencoder is an unsupervised technique for representation learning using neural networks \cite{b19} based on a signal reconstruction problem. \txtc{The output of the neural network is trained to be an approximation of the input, as restricted by the network model itself}. The corresponding optimization problem for applying an autoencoder for our kernel method is defined by,
\beq
\hat{\tht}=\arg\min_{\tht} ||\x-\phi(\tht;\x)||^2,
\label{eq-ae}
\eeq
where both the input $\z$ and output $\ov$ in Eq. (\ref{eq-sdl}) are now set to the x-ray CT image $\x$. The optimization essentially seeks an adaptive CNN representation of the image $\x$, without requiring a large training database. 

\txtc{We specifically consider two types of CNN for $\phi$, as shown in Fig. \ref{fig-1}}. One is the residual encoder-decoder convolutional neural network (RED-CNN) that has been used for low-dose CT denoising \cite{b20}. The other one is a Unet model that is widely used for image segmentation and reconstruction (e.g. \cite{b6}). In the RED-CNN, the encoders consist of a series of 5$\times$5 convolutions followed
by rectified linear unit (ReLU) activation step by step. A series of 5$\times$5 deconvolution layers are used with residual mapping to recover the structural details. For the Unet, a modified structure \cite{b6} is used, which consists of a left-side encoder path and a  right-side decoder path with residual connection from the left to the right. All convolutional filters have a size of 3$\times$3.  Stride 2$\times$2 is used for downsampling and 2$\times$2 bilinear interpolation is used for upsampling. \txtc{This results in approximately $7\times 10^5$ parameters in the RED-CNN model and $3.4\times 10^5$ parameters in the Unet model, as compared to the input image of 32400 pixels in our study.}

Once the model is trained using Eq. (\ref{eq-ae}), the feature set of pixel $j$ is obtained by,
\beq
\f_j^\cnn = \left[\bm{\mathcal{F}}_{\ell}(\hat{\tht};\x)\right]_j,
\eeq
where $\bm{\mathcal{F}}_{\ell}$ denotes the output of the $\ell$th layer of the CNN model, \txtc{which corresponds to the $\ell$th 3D block in Fig. 1. Note that $\ell$ can be only set to a layer that provides multi-channel feature maps of the same size as the input image. Thus, a general choice is the penultimate layer which commonly provides an intermediate output that has the same image size as the final output and also consists of a number of channels. Compared to the intensity-based features, the CNN-derived features form a different nonlinearly transformed space based on which the built kernel method may have a better performance, as illustrated later in the Result section (Fig. \ref{fig-tsne}).}

For the kernel methods, the extracted CNN feature set is fed into the kernel calculation, for instance, using a radial Gaussian kernel,
\beq
\kappa(\f_j^\cnn, \f_l^\cnn) = \exp \left( -\left \|\f_j^\cnn - \f_l^\cnn \right\|^2/2\sigma^2\right),
\eeq
where $\sigma$ is a hyper-parameter which can be set to 1 if the feature set is \txtc{normalized \cite{Wang2015}. 
}
Note that the full consideration of all ($j,l$) pairs for kernel representation would result in a full matrix $\K$, which is impractical for efficient implementation because of its large size. Similar to the previous kernel methods \cite{Wang2015}, $\K$ is built to be sparse using the k-nearest neighbors (kNN) strategy \cite{b17}.

\section{Computer Simulation Studies}

\subsection{Simulation Setup}

We simulated a GE Discovery 690 PET/CT scanner in 2D. The TOF timing resolution of
this PET scanner is about 550 ps. The simulation was conducted using one chest slice of the XCAT phantom. 
The true PET activity image and 511 keV attenuation image are shown in Fig. \ref{fig-pht-xcat}(a) and (b), respectively. The images were first forward projected to generate noise-free sinogram of 11 TOF bins. A 40\% uniform background was included to simulate random and scattered events. Poisson noise was then generated using 5 million expected events. The x-ray CT image at a low-energy 80 keV was also simulated from XCAT and is shown in Fig. \ref{fig-pht-xcat}(c). 

\subsection{Reconstruction Methods for Comparison}

Four types of reconstruction were compared, including (1) standard MLAA \cite{b8}, (2) existing kernel MLAA \cite{Wang20} with $\f_j$ being the pixel intensities of x-ray CT image $\x$ in a 3$\times$3 image patch centered at pixel $j$, (3) RED kernel MLAA: proposed autoencoder kernel method with $\f_j$ extracted from RED-CNN, and (4) Unet kernel MLAA: proposed autoencoder kernel method with $\f_j$ extracted using the Unet. Based on the findings in \cite{Wang20}, we used the 511 keV attenuation map converted from the x-ray CT image as the initial estimate of $\muv$ in the MLAA reconstructions for accelerated convergence. \txtc{All different kernel matrices were built using kNN\cite{b17} with k=50 based on the Euclidean distance between $\f_l$ and $\f_j$ in the same way as used in \cite{Wang2015}}.  All MLAA reconstructions were run for 3000 iterations for the purpose of studying the convergence of different MLAA algorithms. Within each outer iteration,  one inner iteration was used for the $\lam$-estimation step and \txtr{one inner iteration was used for the $\muv$-estimation step. These parameters were chosen for approximately optimal image quality.} 

\subsection{Autoencoder CNN Learning and Feature Visualization}

We used the Adam optimization algorithm to train the RED-CNN and Unet of the x-ray CT image with 300 epochs on a workstation with a Nvidia GeForce RTX 2080 Ti GPU. The learning rates of RED-CNN and Unet were chosen to be $10^{-4}$ and $10^{-2}$, respectively, for approximately optimal performance. \txtc{The training time was approximately 20 seconds for both RED-CNN and Unet in this 2D simulation study. }

To demonstrate the differences between the autoencoder kernel and standard kernel, Fig. \ref{fig-10} visualizes the maps of the feature set used by the two types of kernels. Each subimage corresponds to one element of the feature vector $\f_j$ or $\f_j^\cnn$ at all different pixels. The standard intensity-based kernel was formed from $3\times 3$ neighboring patches, which explains why the feature maps look similar. The Unet-derived CNN features were learned using \txtc{an} autoencoder as the output of the \txtc{penultimate} layer (12 channels) and extracted intrinsic, differential features from the x-ray CT image. 

\txtc{Fig. \ref{fig-tsne} shows the two-dimensional manifold visualization of these high-dimensional (9 and 12, respectively) feature vectors using the t-distributed stochastic neighbor embedding (t-SNE) algorithm \cite{vdM2008}.  \txtr {The t-SNE is a nonlinear dimension reduction technique that locates each high-dimensional data point in a low-dimensional embedding space with intrinsic structures  preserved.} Image pixels (equivalently points in the Fig. \ref{fig-tsne}) are clustered but less organized in the embedding space of the intensity-based features. In comparison, pixels are more organized in the embedding space of the Unet-derived features and follow a continuous change of color, which may be an indication of improvement for kernel MLAA.}

\subsection{Evaluation Metrics}

Different \txtc{MLAA} methods were first compared for the image quality of GCT using the mean squared error (MSE) defined by
\beq
\rm{MSE}(\hat{\muv}) = 10\log_{10}\left(|| \hat{\muv} - \muv^{\rm{true}}||^2 / ||\muv^{\rm{true}} ||^2\right)~~~~\rm(dB),
\eeq
where $\hat{\muv}$ represents the reconstructed GCT image by each MLAA method and $\muv^{\rm{true}}$ denotes the ground truth.
The ensemble bias and standard deviation (SD) of the mean intensity in regions of interest (ROIs) were also calculated to evaluate ROI quantification in a liver region and a bone region shown in Fig. \ref{fig-pht-xcat}(d),
\beq
\textup{Bias} = \frac{1}{c^{\rm{true}}}\left|\overline{c} - c^{\rm{true}} \right|,
\quad \textup{SD} = \frac{1}{c^{\rm{true}}}\sqrt{\frac{1}{\nr- 1}\sum_{i=1}^{\nr}\left| c_i - \txtc{\overline{c}}\right|^2},
\eeq
where $c^{\rm{true}}$ is the noise-free intensity and $\overline{c} = \frac{1}{\nr}\sum_{i=1}^{\nr}c_i$ denotes the mean of $\nr$ realizations. \txtr{$\nr = 10$ in this study. } In addition to the evaluation of bias and SD for ROI quantification,  \txtr{pixel-based ensemble bias and SD in percentage were also calculated and reported in average for specific regions in the same way as used in \txtr{\cite{Wang2015}}.}

Different MLAA algorithms were further compared for dual-energy CT multi-material decomposition. Similarly, image MSE, ROI-based bias and SD, as well as pixel-based bias and SD were calculated for each of the material basis fraction images. Because our focus is on dual-energy CT imaging, we did not intend to evaluate PET activity image reconstruction in this study.

\begin{figure}[t]
	\vspace{-8pt}
	\centering
	{\includegraphics[trim=0.5cm 0cm 0.7cm 0.7cm, clip, width=8cm]{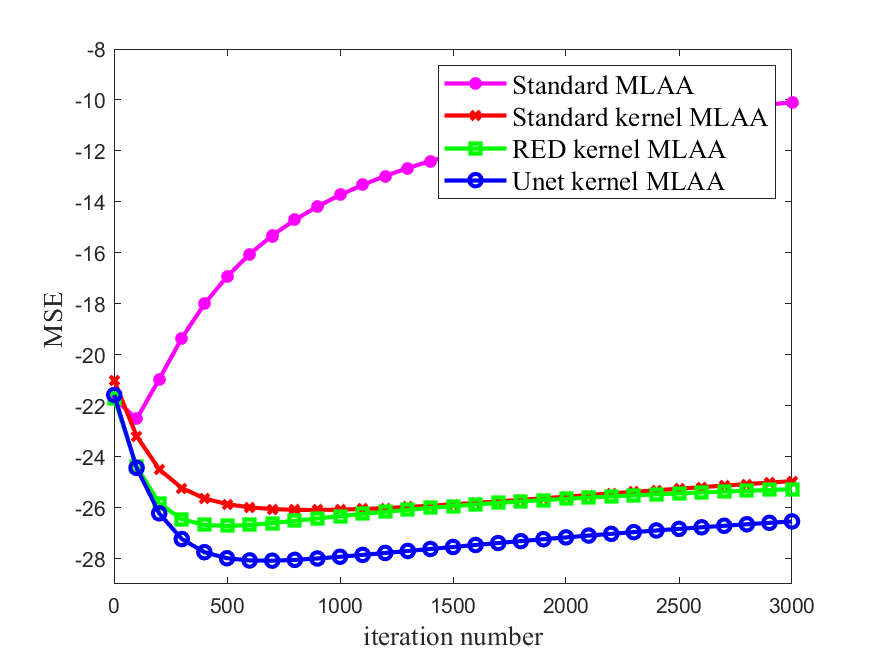}}
	\caption{Plot of image MSE as a function of iteration number for different MLAA reconstruction algorithms.}
	\label{fig-3}
\end{figure}

\begin{figure*}[t]
	\vspace{-8pt}
	\centering
	\subfloat[]{\includegraphics[trim=0.3cm 0cm 0.7cm 0.6cm, clip, width=7.5cm]{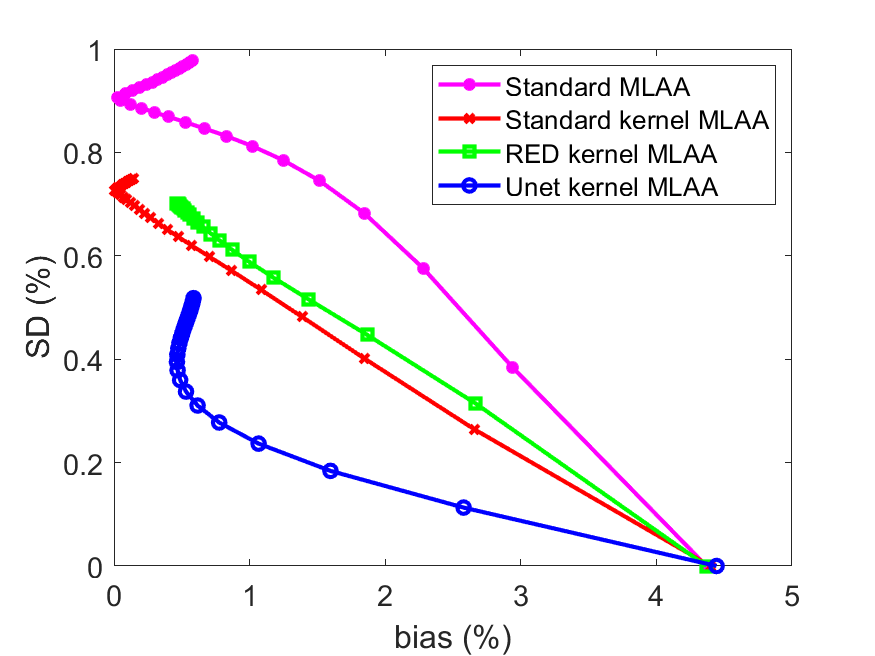}}
	\subfloat[]{\includegraphics[trim=0.3cm 0cm 0.7cm 0.6cm, clip, width=7.5cm]{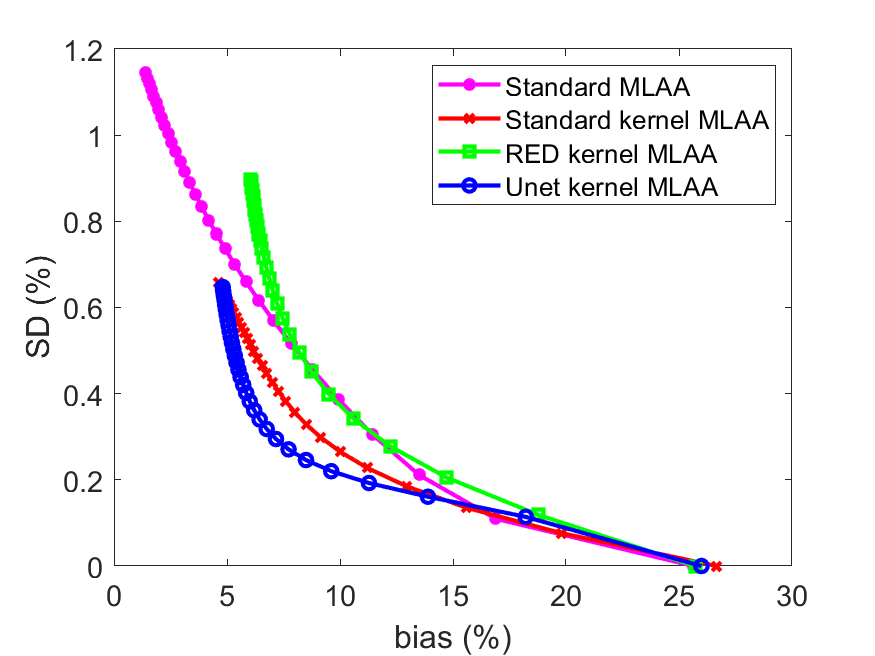}}
	\caption{Plot of bias versus standard deviation (SD) trade-off for GCT ROI quantification. (a) Results of the liver ROI quantification; (b) results of the bone ROI quantification.}
	\label{fig-5}\vspace{-0pt}
\end{figure*}

\begin{figure*}[t]
	\vspace{-15pt}
	\centering
	\subfloat[]{\includegraphics[trim=0.3cm 0cm 0.7cm 0.6cm, clip, width=7.5cm]{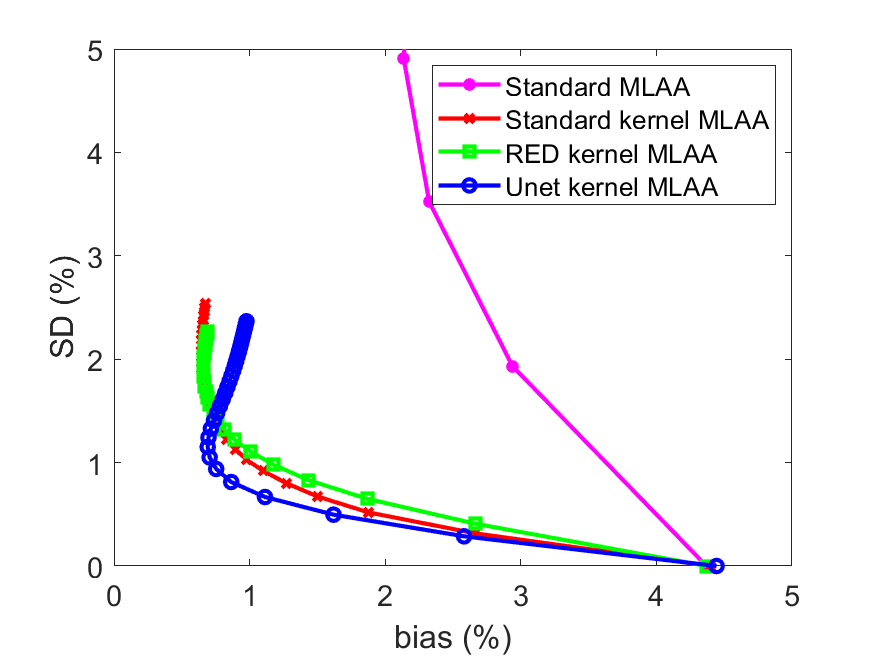}}
	\subfloat[]{\includegraphics[trim=0.3cm 0cm 0.7cm 0.6cm, clip, width=7.5cm]{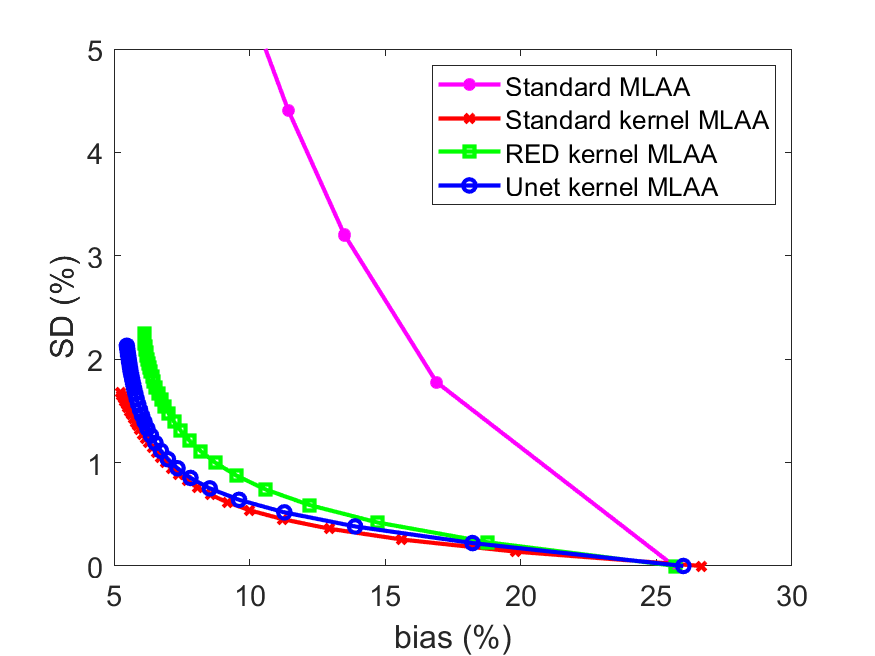}}
	\caption{Plot of pixel-based bias versus standard deviation (SD) trade-off for GCT image evaluation in a liver region (a) and a bone region (b).  For the standard MLAA, only early iterations stay in the display window.}
	\label{fig-13}\vspace{-0pt}
\end{figure*}
\subsection{Comparison Results for GCT Image Quality}

Fig. \ref{fig-4} shows examples of the reconstructed GCT image $\muv$ by different MLAA algorithms with a specific iteration number 600. While the standard MLAA reconstruction was noisy, all the kernel MLAA reconstructions significantly improved the result according to both visual quality and image MSE. The RED kernel MLAA had a slightly better MSE than the standard kernel MLAA and both were with artifacts in the images. In contrast, the Unet kernel MLAA demonstrated least artifacts with good visual quality and achieved the lowest MSE among different algorithms. 

Fig. \ref{fig-3} further shows image MSE as a function of iteration number for different algorithms. The iteration number varies from 0 to 3000 with a step of 100 iterations. Compared to the standard kernel MLAA, the RED kernel MLAA only had a slightly better image MSE at early iterations. The Unet kernel MLAA outperformed the standard kernel MLAA across all iterations. 

The comparison of ensemble bias versus SD for GCT ROI quantification is shown in Fig. \ref{fig-5} by varying the iteration number. As iteration number increases, the bias of ROI quantification is reduced while the SD is increased. All the kernel MLAA algorithms outperformed the standard MLAA.  \txtr{While the RED kernel MLAA did not outperform the standard kernel MLAA, the Unet kernel MLAA achieved the best performance among different algorithms for both liver and bone ROIs. At a fixed bias level, the Unet kernel MLAA has lower SD than the other two kernel-based approaches.  }

Fig. \ref{fig-13} shows the comparison using pixel-based bias versus SD trade-off in the two regions (liver and bone). Again, all the kernel MLAA algorithms outperformed the standard MLAA. The SD level in the pixel-based evaluation is much higher than in the ROI-based evaluation (Fig. \ref{fig-5}) because spatial correlation is taken into account in the latter.  \txtr{The Unet kernel MLAA achieved a better result than the standard kernel MLAA in the liver region for early iterations but became worse for late iterations.  In the bone region, the Unet kernel demonstrated a slightly worse performance than the standard kernel but overall the two algorithms were comparable to each other for the pixel-based bias-SD evaluation in this region. }

\begin{figure*}[t]
	\vspace{-10pt}
	\centering
	
	\subfloat[]{
		\begin{minipage}[t]{0.22\linewidth}
			\centering
			\includegraphics[trim=2.5cm 0.5cm 3.2cm 0cm, clip, height=4.3cm]{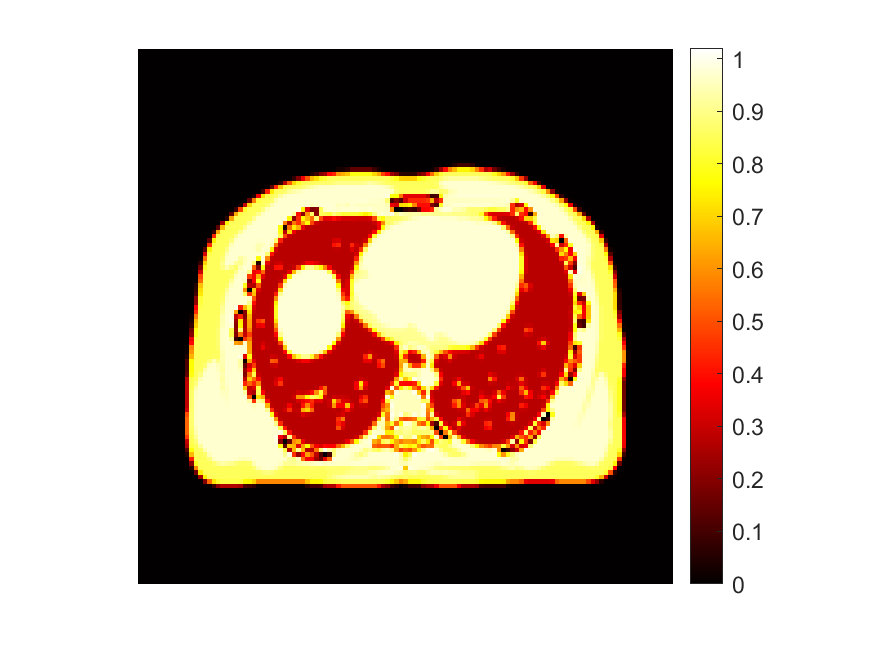}\\
			\includegraphics[trim=2.5cm 0.5cm 3.2cm 0cm, clip, height=4.3cm]{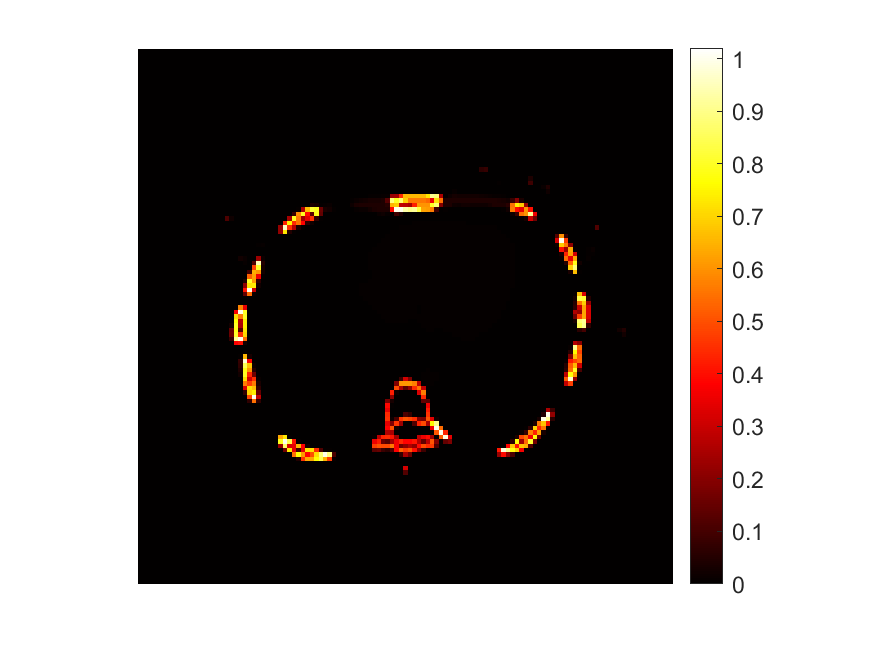}\\
		\end{minipage}%
	}%
	\subfloat[]{
		\begin{minipage}[t]{0.22\linewidth}
			\centering
			\includegraphics[trim=2.5cm 0.5cm 3.2cm 0cm, clip, height=4.3cm]{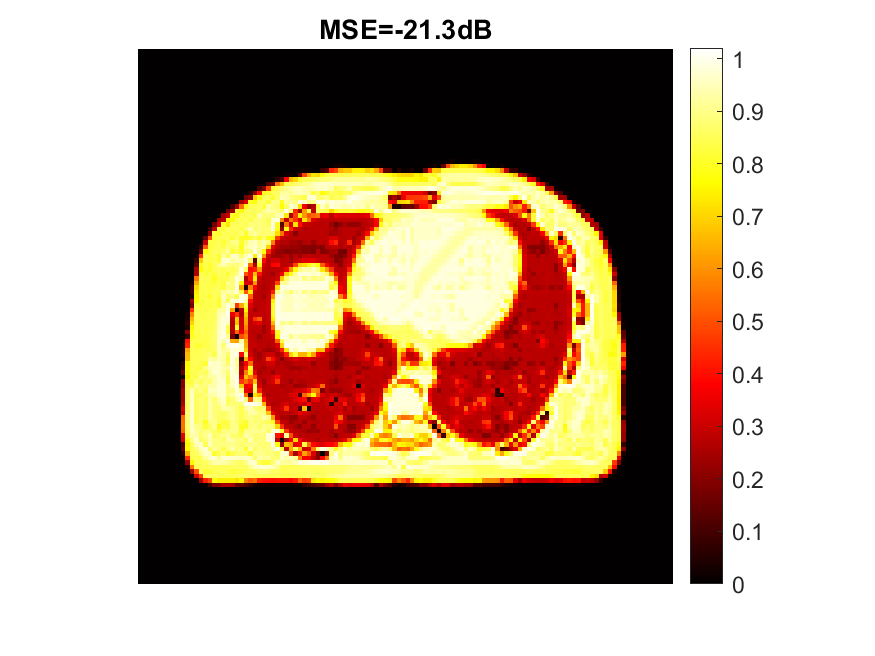}\\
			\includegraphics[trim=2.5cm 0.5cm 3.2cm 0cm, clip, height=4.3cm]{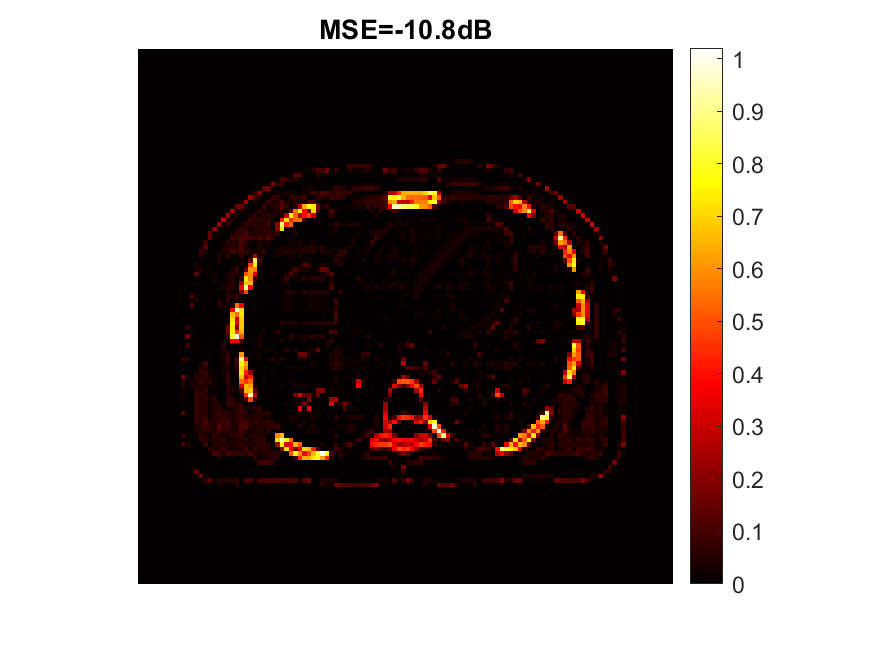}\\
		\end{minipage}%
	}%
	\subfloat[]{
		\begin{minipage}[t]{0.22\linewidth}
			\centering
			\includegraphics[trim=2.5cm 0.5cm 3.2cm 0cm, clip, height=4.3cm]{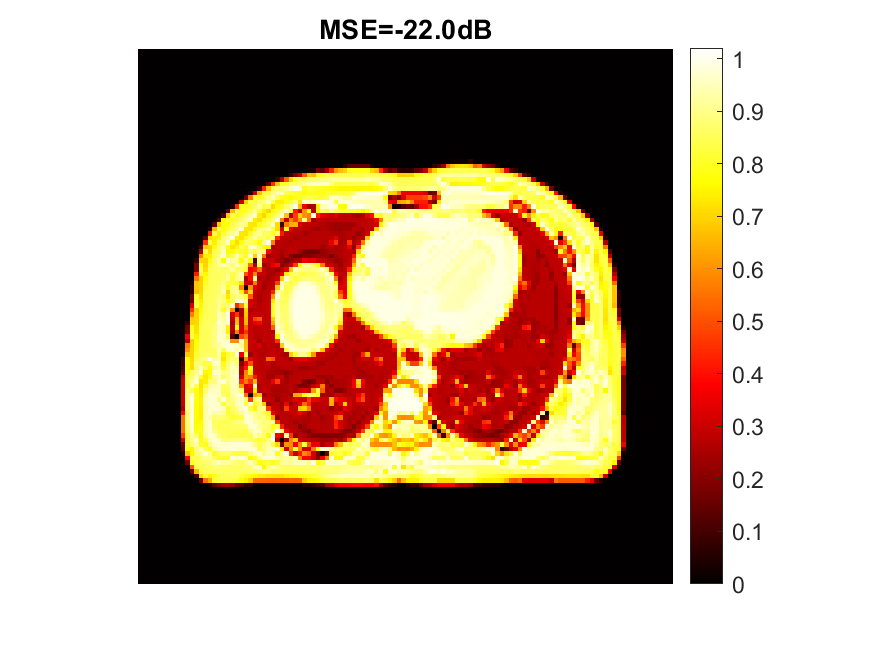}\\
			\includegraphics[trim=2.5cm 0.5cm 3.2cm 0cm, clip, height=4.3cm]{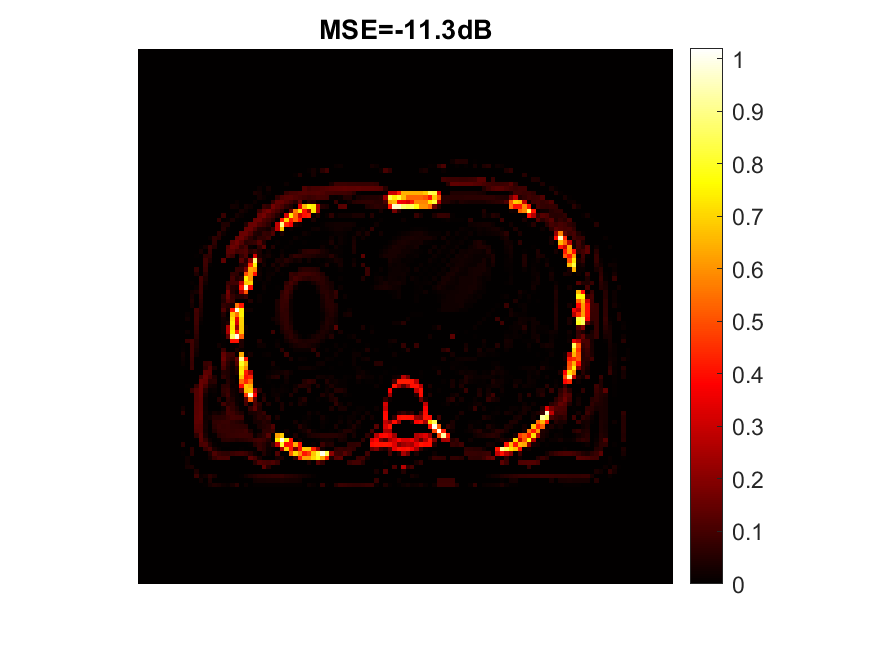}\\
		\end{minipage}%
	}%
	\subfloat[]{
		\begin{minipage}[t]{0.22\linewidth}
			\centering
			\includegraphics[trim=2.5cm 0.5cm 1.5cm 0cm, clip, height=4.3cm]{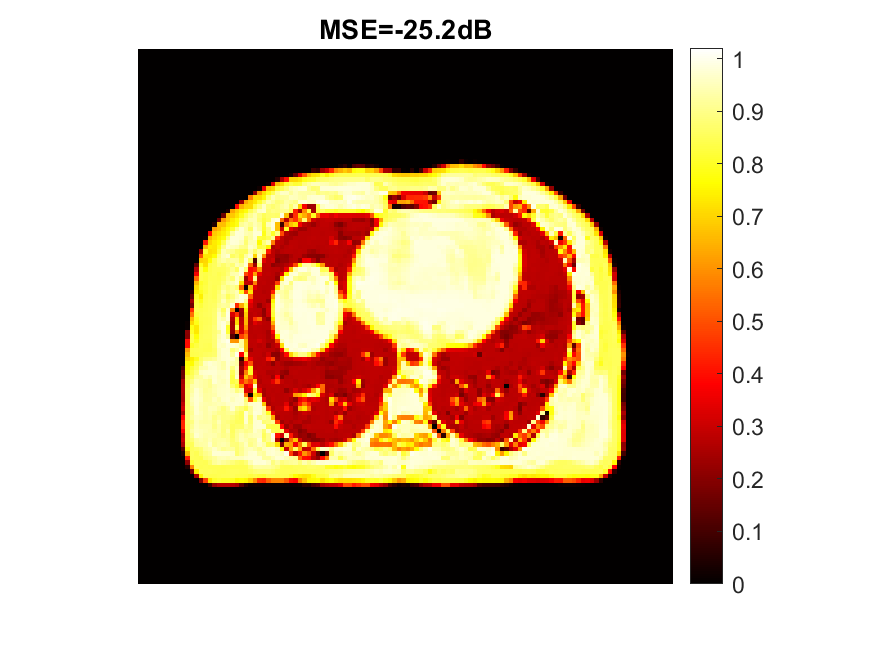}\\
			\includegraphics[trim=2.5cm 0.5cm 1.5cm 0cm, clip, height=4.3cm]{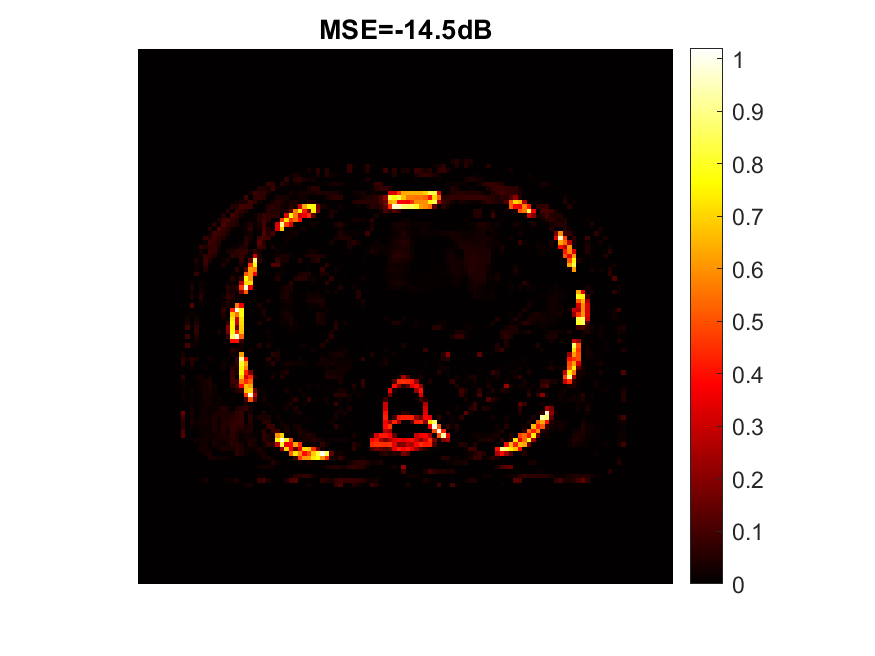}\\
		\end{minipage}%
	}%
	\caption{True and estimated fractional images of two basis materials using different reconstruction algorithms: soft tissue (top row) and bone (bottom row). (a) Ground truth,(b) Standard kernel MLAA, (c) Proposed RED kernel MLAA, (d) Proposed Unet kernel MLAA.}
	\label{fig-7}\vspace{-0pt}
\end{figure*}

\begin{figure*}[t]
	\vspace{-10pt}
	\centering
	\subfloat[Soft tissue]{\includegraphics[trim=0.5cm 0cm 0.5cm 0.5cm, clip, width=7.5cm]{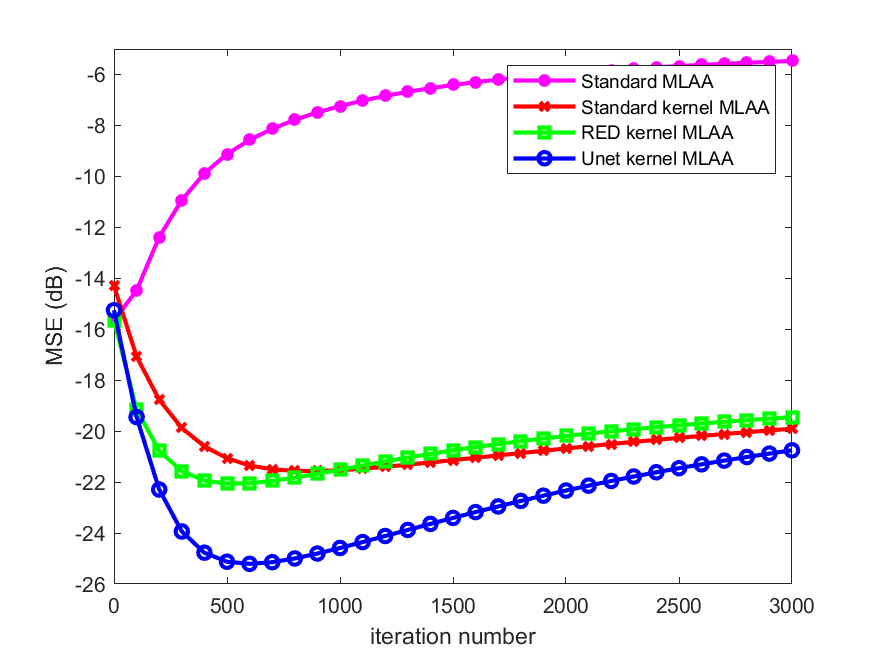}}
	\subfloat[Bone]{\includegraphics[trim=0.5cm 0cm 0.5cm 0.5cm, clip, width=7.5cm]{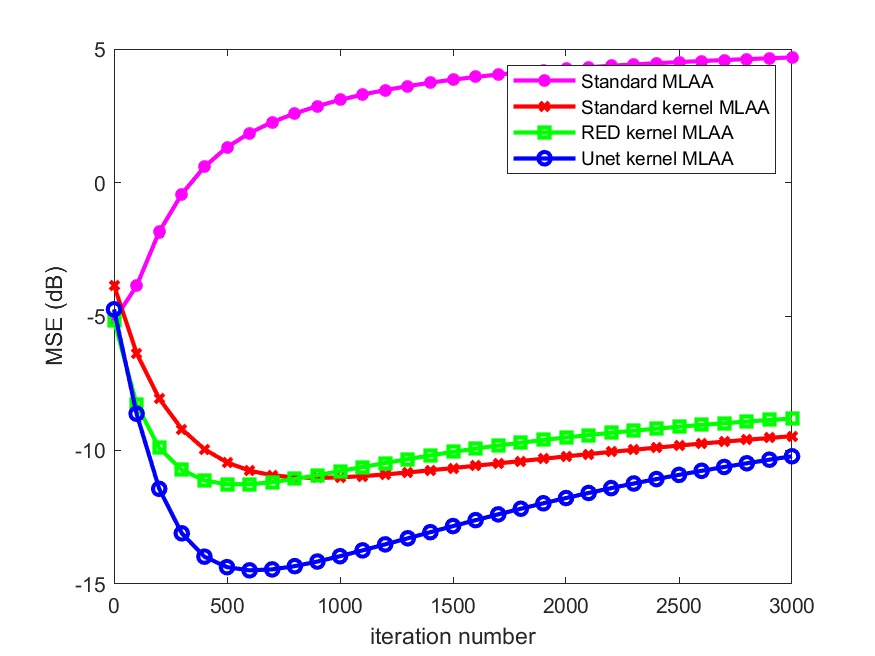}}
	\caption{Plot of image MSE as a function of iteration number for each basis fractional image.}
	\label{fig-6}\vspace{-0pt}
\end{figure*}

\subsection{Comparison Results for Material Decomposition}

Fig. \ref{fig-7} shows the fractional basis images of soft tissue and bone from multi-material decomposition of the PET-enabled dual-energy CT images. The results were obtained from the MLAA reconstructions with 600 iterations. The ground truth of the soft tissue and bone bases was generated using the noise-free data. Compared to the standard MLAA and RED kernel MLAA, the Unet kernel MLAA reconstruction led to better visual quality and decreased image MSE.
Fig. \ref{fig-6} shows image MSE as a function of iteration number for each basis fractional image. All the kernel MLAAs were superior to the standard MLAA, with the best MSE performance from the Unet kernel MLAA.

\begin{figure*}[t]
	\vspace{-10pt}
	\centering
	\subfloat[Soft tissue]{\includegraphics[trim=0cm 0cm 0cm 0.5cm, clip, width=7.5cm]{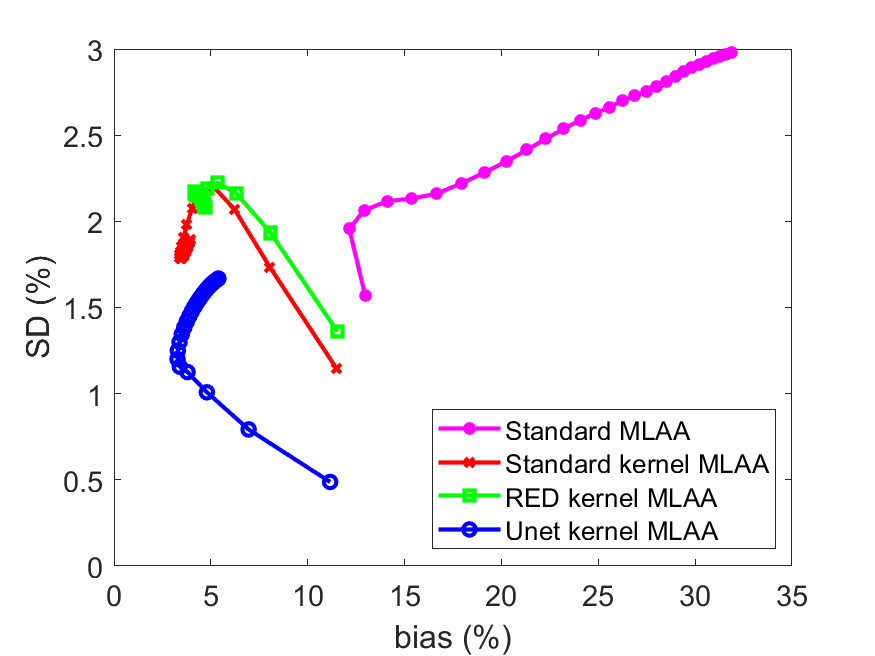}}
	\subfloat[Bone]{\includegraphics[trim=0cm 0cm 0cm 0.5cm, clip, width=7.5cm]{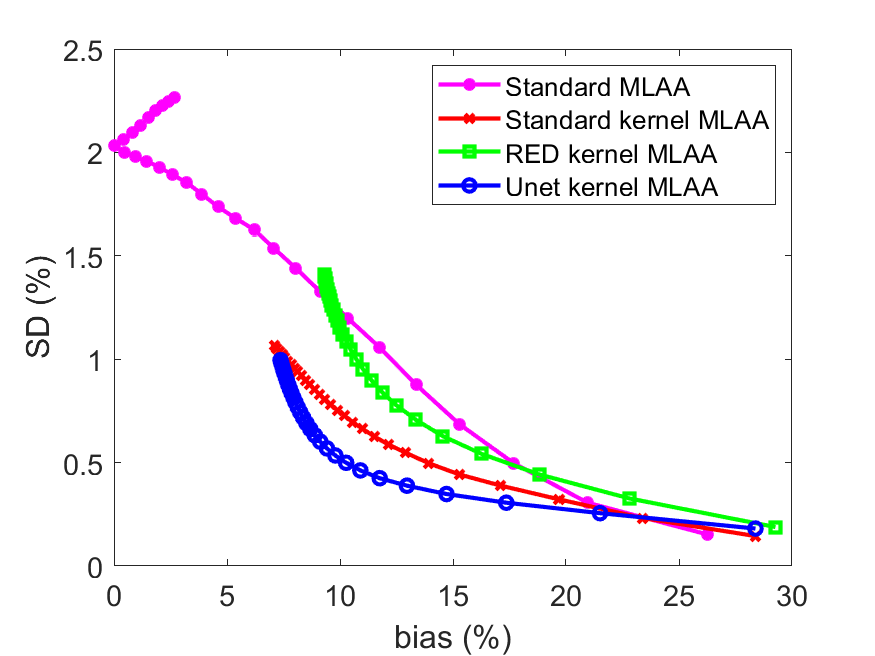}}
	\caption{Bias versus standard deviation trade-off for ROI quantification on the fractional image of (a) soft tissue and (b) bone basis materials.}
	\label{fig-8}
\end{figure*}

To demonstrate the performance of different reconstruction algorithms for ROI quantification on basis fractional images, Fig. \ref{fig-8} shows the bias versus SD trade-off plot for ROI quantification on the soft tissue and bone fractional images. The Unet kernel MLAA outperformed other algorithms for ROI quantification in the liver and bone regions. \txtr{For bone ROI quantification, all kernel MLAAs demonstrated a noticeable bias as compared to the standard kernel MLAA.} The bias in the bone ROI quantification of the fractional basis image was supposed to be propagated from the GCT image reconstruction (as demonstrated in Fig. \ref{fig-5}(b)). 

\begin{figure*}[t]
	\vspace{-12pt}
	\centering
	\subfloat[Soft tissue]{\includegraphics[trim=0cm 0cm 0cm 0.5cm, clip, width=7.5cm]{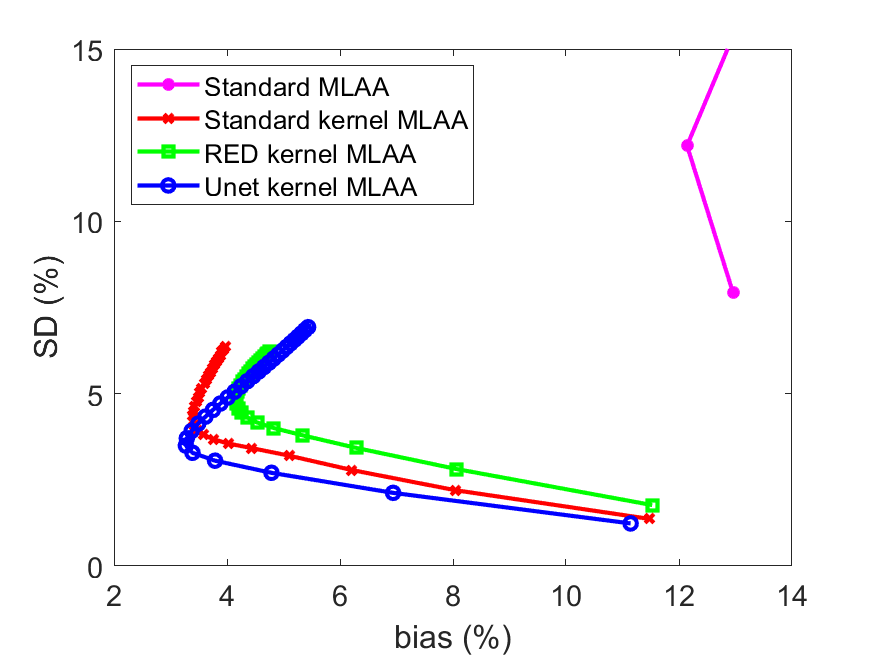}}
	\subfloat[Bone]{\includegraphics[trim=0cm 0cm 0cm 0.5cm, clip, width=7.5cm]{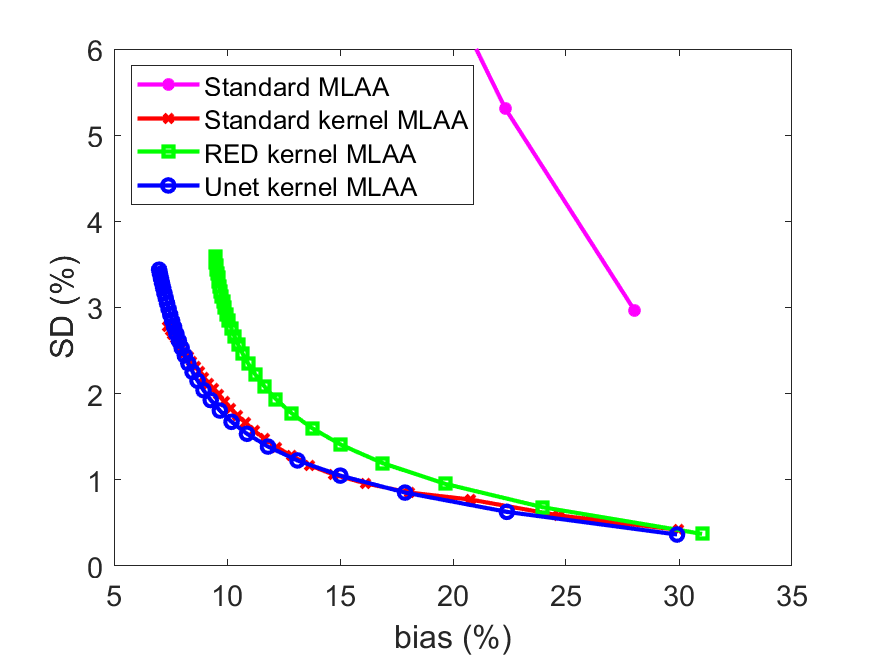}}
	\caption{Pixel-based evaluation of ensemble bias versus standard deviation trade-off for the fractional image of (a) soft tissue and (b) bone basis materials.  For the standard MLAA, only early iterations stay in the display window.}
	\label{fig-14}
\end{figure*}

The result from pixel-based evaluations is shown in Fig. \ref{fig-14}.  \txtr{Overall, the Unet kernel MLAA demonstrated a better or at least comparable performance as compared to the standard kernel MLAA and other algorithms. }

\section{Discussions}

\txtc{This paper combines autoencoder-based representation learning with kernel MLAA to improve PET-enabled dual-energy CT imaging. This work falls into the scope of applying deep learning to MLAA. Compared to direct application of CNN for MLAA (e.g. \cite{Hwang2018, Hwang2019}), the proposed approach has the advantages of not requiring a large training dataset and being patient specific. The training is also computationally efficient as it bases on a single image.}

\txtc{The autoencoder-derived feature maps in Fig. \ref{fig-10}(b) indicate a smoothing effect exists as compared to the intensity-based feature maps in Fig. \ref{fig-10}(a). The smoothing might be due to the downsampling and upsampling in the Unet. In our investigation, applying smoothing to the x-ray CT led to reduced GCT image quality (results not shown), which can be explained by the loss of spatial resolution. This suggests the smoothing effect in the feature maps may have compromised the performance of the proposed approach and possibly caused the bias in the bone region. This problem can be potentially overcome by using a modified Unet architecture (e.g. \cite{Ye2018}) and/or by our other ongoing effort \cite{Li2020} that combines deep image prior \cite{b2, b6} as an implicit regularization on $\alp$ in the kernel MLAA method.}

\txtc{In addition, there are multiple options of layer location to extract the CNN features after the autoencoder learning. This paper does not intend to identify the optimal location but introduces and demonstrates the feasibility for kernel MLAA. It is possible to combine the feature extraction process and kernel optimization into an integrated deep learning framework to further improve the autoencoder kernel MLAA approach. }

\txtc{Based on the quantitative MSE and bias-SD curves shown in this paper, 600 iterations of reconstruction seem a reasonable choice for the kernel MLAA reconstructions. Here the number of iterations is based on not using any ordered subsets. If 30 subsets were used, the needed number of iterations would be approximately 20, which is computationally feasible for clinical practice. }

\txtc{TOF PET data can determine the GCT image but the solution is non-unique due to a scaling constant in the sinogram domain \cite{b9}. The scaling problem was less severe in our application partly because the available x-ray CT provides a very good initial estimate and also the anatomical prior information from x-ray CT could further mitigate the problem in the kernel MLAA reconstructions. Dedicated approaches for the scaling problem are being developed in the field (e.g. \cite{LMK2020}), which may be combined with the kernel MLAA method to improve GCT quantification.}

\txtc{Another limitation of this work is the simulation study used the XCAT phantom that was generated mono-energetically without artifacts. The resulting x-ray CT prior could be very strong and may be less realistic in practice. Our future work will test the method directly using physical phantom and patient data.}

\section{Conclusion}

We have developed an autoencoder kernel MLAA reconstruction method for PET-enabled dual-energy CT. The autoencoder-derived feature set can provide an improved kernel representation to incorporate x-ray CT as the image prior for GCT image reconstruction from PET emission data. Computer simulation results have demonstrated the improvement of the autoencoder kernel MLAA over existing MLAA algorithms for GCT image quality and  dual-energy CT material decomposition. The proposed method can suppress noise, reduce image artifacts, and improve ROI quantification. A weakness of the kernel methods is they may potentially over-smooth a bone region and induce a bias in bone ROI quantification. Our future work will further optimize the method and identify a solution to overcome the limitation.








\end{document}